\documentclass[12pt,a4paper]{article}
\usepackage{amsmath}
\usepackage{latexsym}
\usepackage{amssymb}
\usepackage{graphicx,color}
\makeatletter
\def\rddots{\mathinner{\mkern1mu\raise\p@%
    \vbox{\kern7\p@\hbox{.}}\mkern2mu%
    \raise4\p@\hbox{.}\mkern2mu\raise7\p@\hbox{.}\mkern1mu}}
\makeatother
\makeatletter
\def\eqnarray{%
\stepcounter{equation}%
\let\@currentlabel=\theequation
\global\@eqnswtrue
\global\@eqcnt\z@
\tabskip\@centering
\let\\=\@eqncr
$$\halign to \displaywidth\bgroup\@eqnsel\hskip\@centering
$\displaystyle\tabskip\z@{##}$&\global\@eqcnt\@ne
\hfil$\displaystyle{{}##{}}$\hfil
&\global\@eqcnt\tw@$\displaystyle\tabskip\z@{##}$\hfil
\tabskip\@centering&\llap{##}\tabskip\z@\cr}
\makeatother
\newcommand{\ket}[1]{{\vert{#1}\rangle}}

\newcommand{\fukuso}{{\mathbf C}}

\begin{document}

\title{\sl Algebraic Structure of a Master Equation\\ 
with Generalized Lindblad Form}
\author{
  Kazuyuki FUJII
  \thanks{E-mail address : fujii@yokohama-cu.ac.jp }\\
  ${}^{\dagger}$Department of Mathematical Sciences\\
  Yokohama City University\\
  Yokohama, 236--0027\\
  Japan
  }
\date{}
\maketitle
%
%
%
%
\begin{abstract}
  The quantum damped harmonic oscillator is described by the 
  master equation with usual Lindblad form. The equation has been 
  solved completely by us in arXiv : 0710.2724 [quant-ph]. To 
  construct the general solution a few facts of representation 
  theory based on the Lie algebra $su(1,1)$ were used. 
  
  In this paper we treat a general model described by a master 
  equation with generalized Lindblad form. Then we examine the 
  algebraic structure related to some Lie algebras and construct 
  the interesting approximate solution.
\end{abstract}
%


%
%
%
%

\vspace{5mm}
Quantum Computation (Computer) is one of main subjects in Quantum 
Physics. To realize it we must overcome severe problems arising from 
Decoherence, so we need to study Quantum Open System to control  
decoherence (if possible).

This paper is a series of \cite{Fujii} and \cite{EFS}, and we study 
dynamics of a quantum open system. 
First we explain our purpose in a short manner. See \cite{BP} as a 
general introduction to this subject.
 
We consider a quantum open system $S$ coupled to the environment $E$. 
Then the total system $S+E$ is described by the Hamiltonian
\[
H_{S+E}=H_{S}\otimes {\bf 1}_{E}+{\bf 1}_{S}\otimes H_{E}+H_{I}
\]
where $H_{S}$, $H_{E}$ are respectively the Hamiltonians of the system and 
environment, and $H_{I}$ is the Hamiltonian of the interaction.

\par \noindent
Then under several assumptions (see \cite{BP}) the reduced dynamics of the 
system (which is not unitary !) is given by the Master Equation
\begin{equation}
\label{eq:master-equation}
\frac{\partial}{\partial t}\rho=-i[H_{S},\rho]-{\cal D}(\rho)
\end{equation}
with the dissipator being the usual Lindblad form
\begin{equation}
\label{eq:dissipator}
{\cal D}(\rho)=\frac{1}{2}\sum_{\{j\}}
\left(A_{j}^{\dagger}A_{j}\rho+\rho A_{j}^{\dagger}A_{j}
-2A_{j}\rho A_{j}^{\dagger}\right).
\end{equation}
Here $\rho\equiv \rho(t)$ is the density operator (or matrix) of the system. 

Similarly, the equation of quantum damped harmonic oscillator (see \cite{BP}, 
Section 3.4.6) is given by
\begin{equation}
\label{eq:quantum damped harmonic oscillator}
\frac{\partial}{\partial t}\rho=-i[\omega a^{\dagger}a,\rho]
-
\frac{\mu}{2}
\left(a^{\dagger}a\rho+\rho a^{\dagger}a-2a\rho a^{\dagger}\right)
-
\frac{\nu}{2}
\left(aa^{\dagger}\rho+\rho aa^{\dagger}-2a^{\dagger}\rho{a}\right),
\end{equation}
where $a$ and $a^{\dagger}$ are the annihilation and creation operators of 
the system (for example, an electro--magnetic field mode in a cavity), and 
$\mu,\ \nu$ are some real constants depending on the system (for example, 
a damping rate of the cavity mode).

Since this is one of fundamental equations in quantum theory it is very 
important to construct the general solution. In \cite{BP} or \cite{WS} 
some methods to construct a solution are presented. However, in \cite{EFS} 
we gave the general solution in {\bf the operator algebra level}. 
This is a very important step.

Our method is as follows : 
we clarified a certain algebraic structure arising from the Lie algebra 
$su(1,1)$ and its representation in the equation 
(\ref{eq:quantum damped harmonic oscillator}) and constructed the 
general solution by use of the (well--known) disentangling formula. 
See for example \cite{KF1} and \cite{KF2}. 
The method is popular in Quantum Optics, while it may be not 
in the field of Quantum Open System.  

In this paper we want to generalize the model in order to examine 
deeper algebraic structures. Namely, we consider a master equation 
with generalized Lindblad form defined by
\begin{eqnarray}
\label{eq:quantum damped harmonic oscillator generalized}
\frac{\partial}{\partial t}\rho
=-i[\omega a^{\dagger}a,\rho]
&&-\frac{\mu}{2}
\left(a^{\dagger}a\rho+\rho a^{\dagger}a-2a\rho a^{\dagger}\right)
-\frac{\nu}{2}
\left(aa^{\dagger}\rho+\rho aa^{\dagger}-2a^{\dagger}\rho{a}\right)
\nonumber \\
&&-\frac{\kappa}{2}\left(a^{2}\rho+\rho a^{2}-2a\rho a\right)
  -\frac{\bar{\kappa}}{2}\left( (a^{\dagger})^{2}\rho+
   \rho (a^{\dagger})^{2}-2a^{\dagger}\rho a^{\dagger}\right)
\end{eqnarray}
where $\kappa$ is a complex constant satisfying the condition 
${\mu}{\nu}\geq |\kappa|^{2}$ which ensures the positivity. 
See for example \cite{ABF}\footnote{In the paper it is called 
the Kossakowski--Lindblad form not the generalized Lindblad one. 
It may be suitable.}. 

\par \noindent
Then we examine an algebraic structure related to the Lie algebras  
$su(1,1)$ and $su(2)$, and construct interesting approximate 
solutions by use of it.

In order to solve the equation we use the method in \cite{Fujii} 
once more. For that we review a matrix representation of $a$ and 
$a^{\dagger}$ on the usual Fock space
\[
{\cal F}=\mbox{Vect}_{\fukuso}\{\ket{0},\ket{1},\ket{2},\ket{3},\cdots \};
\quad \ket{n}=\frac{(a^{\dagger})^{n}}{\sqrt{n!}}\ket{0}
\]
like
\begin{eqnarray}
\label{eq:creation-annihilation}
a&=&\mbox{e}^{i\theta}
\left(
\begin{array}{ccccc}
0 & 1 &          &          &        \\
  & 0 & \sqrt{2} &          &        \\
  &   & 0        & \sqrt{3} &        \\
  &   &          & 0        & \ddots \\
  &   &          &          & \ddots
\end{array}
\right),\
a^{\dagger}=\mbox{e}^{-i\theta}
\left(
\begin{array}{ccccc}
0 &          &          &        &        \\
1 & 0        &          &        &        \\
  & \sqrt{2} & 0        &        &        \\
  &          & \sqrt{3} & 0      &        \\
  &          &          & \ddots & \ddots
\end{array}
\right) 
\\
N&=&a^{\dagger}a=
\left(
\begin{array}{ccccc}
0 &   &   &   &        \\
  & 1 &   &   &        \\
  &   & 2 &   &        \\
  &   &   & 3 &        \\
  &   &   &   & \ddots
\end{array}
\right)
\end{eqnarray}
where $\mbox{e}^{i\theta}$ is some phase. Note that $aa^{\dagger}
=a^{\dagger}a+1=N+1$.

For a matrix $X=(x_{ij})\in M({\cal F})$ 
\[X=
\left(
\begin{array}{cccc}
x_{11} & x_{12} & x_{13} & \cdots  \\
x_{21} & x_{22} & x_{23} & \cdots  \\
x_{31} & x_{32} & x_{33} & \cdots  \\
\vdots & \vdots & \vdots & \ddots
\end{array}
\right)
\]
we correspond to the vector $\widehat{X}\in 
{{\cal F}}^{\mbox{dim}_{\fukuso}{\cal F}}$ as
\begin{equation}
\label{eq:correspondence}
X=(x_{ij})\ \longrightarrow\ 
\widehat{X}=(x_{11},x_{12},x_{13},\cdots;x_{21},x_{22},x_{23},\cdots;
x_{31},x_{32},x_{33},\cdots;\cdots \cdots)^{T}
\end{equation}
where $T$ means the transpose. The following formula
\begin{equation}
\label{eq:well--known formula}
\widehat{AXB}=(A\otimes B^{T})\widehat{X}
\end{equation}
holds for $A,B,X\in M({\cal F})$.

Then (\ref{eq:quantum damped harmonic oscillator generalized}) is 
transformed into
\begin{equation}
\label{eq:tranformed equation}
\frac{\partial}{\partial t}\widehat{\rho}(t)=\widehat{H}\widehat{\rho}(t)
\quad \Longrightarrow\quad 
\widehat{\rho}(t)=e^{t\widehat{H}}\widehat{\rho}(0)
\end{equation}
where
\begin{eqnarray}
\label{eq:tranformed form I}
\widehat{H}
&=&-i\omega(N\otimes {\bf 1}-{\bf 1}\otimes N) \nonumber \\
&&-\frac{\mu}{2}\{N\otimes {\bf 1}+{\bf 1}\otimes N -2a\otimes (a^{\dagger})^{T}\}
-\frac{\nu}{2}\{(N+1)\otimes {\bf 1}+{\bf 1}\otimes (N+1) -2a^{\dagger}\otimes a^{T}\}
\nonumber \\
&&-\frac{\kappa}{2}\{a^{2}\otimes {\bf 1}+{\bf 1}\otimes (a^{2})^{T}
-2a\otimes a^{T}\}
-\frac{\bar{\kappa}}{2}\{(a^{\dagger})^{2}\otimes {\bf 1}+
{\bf 1}\otimes ((a^{\dagger})^{2})^{T}-2a^{\dagger}\otimes (a^{\dagger})^{T}\}.
\end{eqnarray}
Moreover it is rewritten as
\begin{eqnarray}
\label{eq:tranformed form II}
\widehat{H}
&=&\frac{\mu-\nu}{2}{\bf 1}\otimes {\bf 1}
-(\mu+\nu)\frac{N\otimes {\bf 1}+{\bf 1}\otimes N+{\bf 1}\otimes {\bf 1}}{2}
+\nu a^{\dagger}\otimes a^{T}+\mu a\otimes (a^{\dagger})^{T}  \nonumber \\
&&
-2i\omega \frac{N\otimes {\bf 1}-{\bf 1}\otimes N}{2}
+\bar{\kappa}\left\{a^{\dagger}\otimes (a^{\dagger})^{T}-
\frac{(a^{\dagger})^{2}\otimes {\bf 1}+{\bf 1}\otimes ((a^{\dagger})^{2})^{T}}{2}\right\}
\nonumber \\
&&
\qquad \qquad \qquad \qquad \quad \ \ +\kappa\left\{a\otimes a^{T}-
\frac{a^{2}\otimes {\bf 1}+{\bf 1}\otimes (a^{2})^{T}}{2}\right\}.
\end{eqnarray}

Now let us examine the algebraic structure of $\widehat{H}$.

\par \noindent 
By setting
\begin{equation}
\label{eq:K's I}
\tilde{K}_{3}=\frac{1}{2}(N\otimes {\bf 1}+{\bf 1}\otimes N+
{\bf 1}\otimes {\bf 1}),\quad
\tilde{K}_{+}=a^{\dagger}\otimes a^{T},\quad
\tilde{K}_{-}=a\otimes (a^{\dagger})^{T}
\end{equation}
where $N^{T}=N$, then we have
\begin{equation}
\label{eq:K's relation I}
[\tilde{K}_{3},\tilde{K}_{+}]=\tilde{K}_{+},\quad 
[\tilde{K}_{3},\tilde{K}_{-}]=-\tilde{K}_{-},\quad 
[\tilde{K}_{+},\tilde{K}_{-}]=-2\tilde{K}_{3}.
\end{equation}
Namely, $\{\tilde{K}_{3},\tilde{K}_{+},\tilde{K}_{-}\}$ is a set of 
generators of $su(1,1)$ algebra, \cite{EFS}.

\par \noindent 
By setting
\begin{equation}
\label{eq:J's}
J_{3}=\frac{1}{2}(N\otimes {\bf 1}-{\bf 1}\otimes N),\quad
J_{+}=a^{\dagger}\otimes (a^{\dagger})^{T},\quad
J_{-}=a\otimes a^{T}
\end{equation}
, then we have
\begin{equation}
\label{eq:J's relation}
[J_{3},J_{+}]=J_{+},\quad [J_{3},J_{-}]=-J_{-},\quad 
[J_{+},J_{-}]=2J_{3}.
\end{equation}
Namely, $\{J_{3},J_{+},J_{-}\}$ is a set of generators of 
$su(2)$ algebra.

\par \noindent 
By setting
\begin{equation}
\label{eq:K's}
K_{3}=\frac{1}{2}(N\otimes {\bf 1}-{\bf 1}\otimes N),\
K_{+}=\frac{1}{2}\left\{(a^{\dagger})^{2}\otimes {\bf 1}+
{\bf 1}\otimes ((a^{\dagger})^{2})^{T}\right\},\
K_{-}=\frac{1}{2}\left\{a^{2}\otimes {\bf 1}+
{\bf 1}\otimes (a^{2})^{T}\right\}
\end{equation}
, then we have
\begin{equation}
\label{eq:K's relation}
[K_{3},K_{+}]=K_{+},\quad [K_{3},K_{-}]=-K_{-},\quad 
[K_{+},K_{-}]=-2K_{3}.
\end{equation}
Namely, $\{K_{3},K_{+},K_{-}\}$ is a set of generators of 
$su(1,1)$ algebra.

\par \noindent 
By setting
\begin{equation}
\label{eq:L's}
L_{3}=\frac{1}{2}(N\otimes {\bf 1}-{\bf 1}\otimes N),\quad
L_{+}=J_{+}-K_{+},\quad
L_{-}=J_{-}-K_{-}
\end{equation}
, then we have
\begin{equation}
\label{eq:L's relation}
[L_{3},L_{+}]=L_{+},\quad [L_{3},L_{-}]=-L_{-},\quad 
[L_{+},L_{-}]=0.
\end{equation}
We also note that
\begin{equation}
\label{eq:some relation-I}
[J_{+},K_{+}]=[J_{-},K_{-}]=0.
\end{equation}

\par \noindent
However, $\{\tilde{K}_{3},\tilde{K}_{+},\tilde{K}_{-}\}$ 
and $\{L_{3},L_{+},L_{-}\}$ don't commute except for
\begin{equation}
\label{eq:some relation-II}
[L_{3},\tilde{K}_{3}]=[L_{3},\tilde{K}_{+}]=
[L_{3},\tilde{K}_{-}]=0,
\end{equation}
see \cite{EFS}. 

A comment is in order.\ \ For later convenience let us 
write down the remaining commutators. 
\begin{eqnarray*}
&&[\tilde{K}_{3},L_{+}]=
-\frac{1}{2}\{(a^{\dagger})^{2}\otimes {\bf 1}-
{\bf 1}\otimes ((a^{\dagger})^{2})^{T}\},\quad
[\tilde{K}_{3},L_{-}]=
\frac{1}{2}\{(a^{2}\otimes {\bf 1}-
{\bf 1}\otimes (a^{2})^{T}\}, \\
&&[\tilde{K}_{+},L_{+}]=
a^{\dagger}\otimes (a^{\dagger})^{T}-
(a^{\dagger})^{2}\otimes {\bf 1},\quad 
[\tilde{K}_{+},L_{-}]=
a\otimes a^{T}-{\bf 1}\otimes (a^{2})^{T}, \\
&&[\tilde{K}_{-},L_{+}]=
-a^{\dagger}\otimes (a^{\dagger})^{T}+
{\bf 1}\otimes ((a^{\dagger})^{2})^{T},\quad 
[\tilde{K}_{-},L_{-}]=
-a\otimes a^{T}+a^{2}\otimes {\bf 1}.
\end{eqnarray*}

\vspace{5mm}
Then (\ref{eq:tranformed form II}) is rewritten like 
\begin{equation}
\label{eq:tranformed form III}
\widehat{H}
=\frac{\mu-\nu}{2}{\bf 1}\otimes {\bf 1}
-(\mu+\nu)\tilde{K}_{3}+\nu \tilde{K}_{+}+\mu \tilde{K}_{-}
-2i\omega {L}_{3}+\bar{\kappa}{L}_{+}+\kappa{L}_{-}.
\end{equation}
What we want to do is to calculate the evolution operator 
$e^{t\widehat{H}}$, which is in general not easy. 
Since $\{\tilde{K}_{3},\tilde{K}_{+},\tilde{K}_{-}\}$ and 
$\{L_{3},L_{+},L_{-}\}$ don't commute it is reasonable to assume 
\begin{equation}
\label{eq:evolution operator}
e^{t\widehat{H}}\approx e^{\frac{\mu-\nu}{2}t}
e^{t\{-(\mu+\nu)\tilde{K}_{3}+\nu \tilde{K}_{+}+\mu \tilde{K}_{-}\}}
e^{t\{-2i\omega {L}_{3}+\bar{\kappa}{L}_{+}+\kappa{L}_{-}\}}
\end{equation}
as {\bf the first approximation}. 

A comment is in order.\ In place of (\ref{eq:evolution operator}) 
it may be also reasonable to take
\[
e^{t\widehat{H}}\approx e^{\frac{\mu-\nu}{2}t}
e^{t\bar{\kappa}{L}_{+}}
e^{t\{-2i\omega {L}_{3}-(\mu+\nu)\tilde{K}_{3}+
\nu \tilde{K}_{+}+\mu \tilde{K}_{-}\}}
e^{t\kappa{L}_{-}}.
\]
However, we don't consider this approximation in the paper.

\vspace{3mm}
By the way, we have calculated the term \ \ 
$
e^{t\{-(\mu+\nu)\tilde{K}_{3}+\nu \tilde{K}_{+}+\mu \tilde{K}_{-}\}}
$
\ \ in \cite{EFS}. The result is
\begin{equation}
\label{eq:disentangling formula}
e^{t\{-(\mu+\nu)\tilde{K}_{3}+\nu \tilde{K}_{+}+\mu \tilde{K}_{-}\}}
=
e^{G(t)K_{+}}e^{-2\log(F(t))K_{3}}e^{E(t)K_{-}},
\end{equation}
or more explicitly
\begin{eqnarray}
\label{eq:disentangling formula II}
\mbox{RHS}
=
&&\frac{1}{F(t)}
\exp\left(G(t)a^{\dagger}\otimes a^{T}\right)
\left(
\exp\left(-\log(F(t))N\right)\otimes 
\exp\left(-\log(F(t))N\right)^{T}
\right)\times  \nonumber \\
&&\qquad \ \exp\left(E(t)a\otimes (a^{\dagger})^{T}\right)
\end{eqnarray}
with
\begin{eqnarray}
E(t)&=&\frac{\frac{2\mu}{\mu-\nu}\sinh\left(\frac{\mu-\nu}{2}t\right)}
     {\cosh\left(\frac{\mu-\nu}{2}t\right)+\frac{\mu+\nu}{\mu-\nu}
      \sinh\left(\frac{\mu-\nu}{2}t\right)},\quad
G(t)=\frac{\frac{2\nu}{\mu-\nu}\sinh\left(\frac{\mu-\nu}{2}t\right)}
     {\cosh\left(\frac{\mu-\nu}{2}t\right)+\frac{\mu+\nu}{\mu-\nu}
      \sinh\left(\frac{\mu-\nu}{2}t\right)}   \nonumber \\
F(t)&=&\cosh\left(\frac{\mu-\nu}{2}t\right)+
     \frac{\mu+\nu}{\mu-\nu}\sinh\left(\frac{\mu-\nu}{2}t\right).
\end{eqnarray}

\vspace{5mm}
On the other hand, we can calculate the term \  
$e^{t\{-2i\omega L_{3}+\bar{\kappa} L_{+}+\kappa L_{-}\}}$ 
easily because of the relation $[L_{+},L_{-}]=0$ in 
(\ref{eq:L's relation}). The result is
\begin{equation}
\label{eq:disentangling formula III}
e^{t\{-2i\omega {L}_{3}+\bar{\kappa}{L}_{+}+\kappa{L}_{-}\}}
=
e^{f(t)L_{+}}e^{g(t)L_{3}}e^{l(t)L_{-}}
=
e^{f(t)(J_{+}-K_{+})}e^{g(t)L_{3}}e^{l(t)(J_{-}-K_{-})}
\end{equation}
with
\begin{equation}
f(t)=\frac{e^{-2i\omega t}-1}{-2i\omega}\bar{\kappa},\quad
g(t)=-2i\omega t,\quad
l(t)=\frac{e^{-2i\omega t}-1}{-2i\omega}\kappa.
\end{equation}
More explicitly
\begin{eqnarray}
\label{eq:disentangling formula IV}
\mbox{RHS}
=
&&
\exp\left(f(t)a^{\dagger}\otimes (a^{\dagger})^{T}\right)
\left(
\exp\left(-\frac{f(t)}{2}(a^{\dagger})^{2}\right)\otimes 
\exp\left(-\frac{f(t)}{2}((a^{\dagger})^{2})^{T}\right)
\right)\times \nonumber \\
&&\exp\left(\frac{g(t)}{2}N\right)\otimes 
\exp\left(-\frac{g(t)}{2}N^{T}\right)\times  \nonumber \\
&&
\exp\left(l(t)a\otimes a^{T}\right)
\left(
\exp\left(-\frac{l(t)}{2}a^{2}\right)\otimes 
\exp\left(-\frac{l(t)}{2}(a^{2})^{T}\right)
\right)
\end{eqnarray}
because of (\ref{eq:some relation-I}).

Therefore our approximate solution is
\begin{eqnarray*}
\widehat{\rho}(t)
&&\approx 
e^{\frac{\mu-\nu}{2}t}
e^{t\{-(\mu+\nu)\tilde{K}_{3}+\nu \tilde{K}_{+}+\mu \tilde{K}_{-}\}}
e^{t\{-2i\omega {L}_{3}+\bar{\kappa}{L}_{+}+\kappa{L}_{-}\}}
\widehat{\rho}(0) \\
&&=
\frac{\mbox{e}^{\frac{\mu-\nu}{2}t}}{F(t)}
\exp\left(G(t)a^{\dagger}\otimes a^{T}\right)\times \\
&&\qquad \quad \ \
\left(\exp\left(-\log(F(t))N\right)\otimes 
\exp\left(-\log(F(t))N\right)^{T}
\right)\times  \\
&&\qquad \quad \ \ \exp\left(E(t)a\otimes (a^{\dagger})^{T}\right) 
\times  \\
&&\qquad \quad \ \
\exp\left(f(t)a^{\dagger}\otimes (a^{\dagger})^{T}\right)
\left(
\exp\left(-\frac{f(t)}{2}(a^{\dagger})^{2}\right)\otimes 
\exp\left(-\frac{f(t)}{2}((a^{\dagger})^{2})^{T}\right)
\right)\times \nonumber \\
&&\qquad \quad \ \
\exp\left(\frac{g(t)}{2}N\right)\otimes 
\exp\left(-\frac{g(t)}{2}N^{T}\right)\times  \nonumber \\
&&\qquad \quad \ \
\exp\left(l(t)a\otimes a^{T}\right)
\left(
\exp\left(-\frac{l(t)}{2}a^{2}\right)\otimes 
\exp\left(-\frac{l(t)}{2}(a^{2})^{T}\right)
\right)
\widehat{\rho}(0)
\end{eqnarray*}
and we restore this form to the usual one by use of 
(\ref{eq:well--known formula}). The result is
\begin{eqnarray}
\label{eq:final form}
\rho(t)=
\frac{\mbox{e}^{\frac{\mu-\nu}{2}t}}{F(t)}
&&\sum_{n=0}^{\infty}
\frac{G(t)^{n}}{n!}(a^{\dagger})^{n}
\{
\exp\left(\{-\log(F(t))\}N\right)\times  \nonumber \\
&&
\left\{
\sum_{m=0}^{\infty}
\frac{E(t)^{m}}{m!}a^{m}\phi(t)(a^{\dagger})^{m}
\right\}
\exp\left(\{-\log(F(t))\}N\right)
\}
a^{n}
\end{eqnarray}
and
\begin{eqnarray}
\label{eq:final form}
\phi(t)=
&&\sum_{k=0}^{\infty}
\frac{f(t)^{k}}{k!}(a^{\dagger})^{k}
\exp\left(-\frac{f(t)}{2}(a^{\dagger})^{2}\right)
\left\{
\exp\left(\frac{g(t)}{2}N\right)\times  
\right.  \nonumber \\
&&
\left.
\left\{
\sum_{j=0}^{\infty}
\frac{l(t)^{j}}{j!}
a^{j}
\exp\left(-\frac{l(t)}{2}a^{2}\right)
\rho(0)
\exp\left(-\frac{l(t)}{2}a^{2}\right)
a^{j}
\right\}\exp\left(-\frac{g(t)}{2}N\right)  
\right\}  \nonumber \\
&&
\quad \exp\left(-\frac{f(t)}{2}(a^{\dagger})^{2}\right)
(a^{\dagger})^{k}.  
\end{eqnarray}
This is indeed complicated.

\vspace{3mm}
To construct the general solution to the equation 
(\ref{eq:quantum damped harmonic oscillator generalized}) is very 
important in not only Physics but also Mathematics. However, it is 
not easy at the moment, so we only constructed some approximate 
solution. In the very near future we would like to do it.

\vspace{5mm}
In this paper we revisited the quantum damped harmonic oscillator 
with generalized Lindblad form and constructed some approximate 
solution in the operator algebra level. 

The model is very important to understand several phenomena related to 
quantum open systems, so the general solution is required. 

On the other hand we are studying some related topics from a different 
point of view, see \cite{KF3} and \cite{SR}.

Lastly, we conclude the paper by stating our motivation. We are studying 
a model of quantum computation (computer) based on Cavity QED (see 
\cite{FHKW1} and \cite{FHKW2}), so in order to construct a more realistic 
model of (robust) quantum computer we have to study severe problems 
coming from decoherence. 

For example, we have to study the quantum damped Jaynes--Cummings model 
(in our terminology) whose phenomenological master equation for the 
density operator is given by
\begin{equation}
\label{eq:quantum damped Jaynes-Cummings}
\frac{\partial}{\partial t}\rho=-i[H_{JC},\rho]
-
\frac{\mu}{2}
\left(a^{\dagger}a\rho+\rho a^{\dagger}a-2a\rho a^{\dagger}\right)
-
\frac{\nu}{2}
\left(aa^{\dagger}\rho+\rho aa^{\dagger}-2a^{\dagger}\rho{a}\right),
\end{equation}
where $H_{JC}$ is the well--known Jaynes-Cummings Hamiltonian given by
\begin{eqnarray}
H_{JC}
&=&
\frac{\omega_{0}}{2}\sigma_{3}\otimes {\bf 1}+ 
\omega_{0}1_{2}\otimes a^{\dagger}a +
\Omega\left(\sigma_{+}\otimes a+\sigma_{-}\otimes a^{\dagger}\right)
\\
&=&
\left(
  \begin{array}{cc}
    \frac{\omega_{0}}{2}+\omega_{0}N & \Omega a             \\
    \Omega a^{\dagger} & -\frac{\omega_{0}}{2}+\omega_{0}N
  \end{array}
\right) \nonumber
\end{eqnarray}
with
\[
\sigma_{+} = 
\left(
  \begin{array}{cc}
    0 & 1 \\
    0 & 0
  \end{array}
\right), \quad 
\sigma_{-} = 
\left(
  \begin{array}{cc}
    0 & 0 \\
    1 & 0
  \end{array}
\right), \quad 
\sigma_{3} = 
\left(
  \begin{array}{cc}
    1 & 0  \\
    0 & -1
  \end{array}
\right), \quad 
{\bf 1}_{2} = 
\left(
  \begin{array}{cc}
    1 & 0 \\
    0 & 1
  \end{array}
\right). 
\]
Note that $\rho \in M(2;\fukuso)\otimes M({\cal F})=
M(2;M({\cal F}))$, where $M({\cal F})$ is the set of all 
operators on the Fock space ${\cal F}$. 
See for example \cite{Scala et al-1}, \cite{Scala et al-2}.

Furthermore, it may be possible to treat the generalized master 
equation given by
\begin{eqnarray}
\label{eq:quantum damped Jaynes-Cummings II}
\frac{\partial}{\partial t}\rho
&=&-i[H_{JC},\rho]
-
\frac{\mu}{2}
\left(a^{\dagger}a\rho+\rho a^{\dagger}a-2a\rho a^{\dagger}\right)
-
\frac{\nu}{2}
\left(aa^{\dagger}\rho+\rho aa^{\dagger}-2a^{\dagger}\rho{a}\right)
\nonumber \\
&&-\frac{\kappa}{2}\left(a^{2}\rho+\rho a^{2}-2a\rho a\right)
  -\frac{\bar{\kappa}}{2}\left( (a^{\dagger})^{2}\rho+
   \rho (a^{\dagger})^{2}-2a^{\dagger}\rho a^{\dagger}\right)
\end{eqnarray}
with the condition ${\mu}{\nu}\geq |\kappa|^{2}$ similarly  
in this paper.

These equations ((\ref{eq:quantum damped Jaynes-Cummings}), 
(\ref{eq:quantum damped Jaynes-Cummings II})) are very hard 
to solve in the operator algebra level,  
so even constructing approximate solutions is not easy. 
This is our future task, \cite{Fujii-more}.


\end{document}